# The origin of insulating and non-ferromagnetic SrRuO$_3$ monolayers

Zeeshan Ali[1], Zhen Wang[1,3], Andrew O'Hara[2], Mohammad Saghayezhian[1], Donghan Shin[2], Yimei Zhu[3], Sokrates T. Pantelides[2,4], and Jiandi Zhang[1,5]

[1]*Department of Physics & Astronomy, Louisiana State University, Baton Rouge, LA 70803, USA.*
[2]*Department of Physics and Astronomy, Vanderbilt University, Nashville, Tennessee 37235, USA.*
[3]*Condensed Matter Physics & Materials Science, Department, Brookhaven National Laboratory, Upton, NY 11973, USA.*
[4]*Department of Electrical Computer Engineering, Vanderbilt University, Nashville, Tennessee 37235, USA.*
[5]*Beijing National Laboratory for Condensed Matter Physics, Institute of Physics, Chinese Academy of Sciences, 100190 Beijing, China.*

**Abstract**

The electro-magnetic properties of ultrathin epitaxial ruthenate films have long been the subject of debate. Here we combine experimental with theoretical investigations of $(SrTiO_3)^5$-$(SrRuO_3)^n$-$(SrTiO_3)^5$ ($STO^5$-$SRO^n$-$STO^5$) heterostructures with $n = 1$ and 2 unit cells, including extensive atomic-resolution scanning-transmission-electron-microscopy imaging, electron-energy-loss-spectroscopy chemical mapping, as well as transport and magneto-transport measurements. The experimental data demonstrate that the $STO^5$-$SRO^2$-$STO^5$ heterostructure is stoichiometric, metallic, and ferromagnetic with $T_C \sim 128$ K, even though it lacks the characteristic bulk-SRO octahedral tilts and matches the cubic STO structure. In contrast, the $STO^5$-$SRO^1$-$STO^5$ heterostructure features Ru-Ti intermixing in the RuO$_2$ layer, also without octahedral tilts, but is accompanied by a loss of metallicity and ferromagnetism. Density-functional-theory calculations show that stoichiometric $n = 1$ and $n = 2$ heterostructures are metallic and ferromagnetic with no octahedral tilts, while non-stoichiometry in the Ru sublattice in the $n = 1$ case opens an energy gap and induces antiferromagnetic ordering. Thus, the results indicate that the observed non-stoichiometry is the cause of the observed loss of metallicity and ferromagnetism in the $n = 1$ case.

## I.   INTRODUCTION

Complex-oxide heterostructures have generated significant interest because of their diverse emergent phenomena, including ferromagnetism [1–3], ferroelectricity [4,5], interfacial 2D electron gas [6], topological spin texture [7], strain-induced superconductivity [8], etc. However, one limitation in the design of thin-film-based oxide heterostructures is the occurrence of dead layers exhibiting insulating and non-ferromagnetic (FM) behavior below a certain film thickness [9–11]. As a prototype example, bulk 4$d$ transition-metal oxide SrRuO$_3$ (SRO) has an orthorhombic perovskite lattice structure [12] and a FM-metallic ground state with a Curie temperature of ~160 K [13,14]. However, ultrathin films of SRO grown on substrates such as SrTiO$_3$ (STO) exhibit intriguing properties that are different from bulk counterparts [15–19], including the occurrence of metal-insulator transition (MIT) and non-FM state [11,20–24].

Historically, the earliest experimental investigation by Toyota et al. [25,26] initiated interest in thickness-dependent properties of SRO, where MIT is observed to arise at a film thickness of 4-5 u.c. Nonetheless, different values have since been reported for the MIT critical thickness, from 2 unit-cells (u.c.) [11], to 3 u.c. [20,21,24], and 4 u.c. [27], with the variance usually attributed to the degree of disorder existing in films. Numerous theoretical studies have explored the origin of MIT, while remaining inconclusive regarding the critical thickness and nature of the ground state, i.e., whether FM or antiferromagnetic (AFM) insulator [28–31]. Several authors suggested that reducing film thickness may enhance electronic correlations [18,32,33] and result in a structural transition [34]. Rondinelli et al. [28], however, reported a comprehensive theoretical investigation of these effects and found that neither enhanced electronic correlations nor structural transitions could reproduce the experimentally observed MIT, leading to a suggestion that extrinsic effects (such as surface disorder and defects) or dynamic spin correlations may be the dominant factor. To overcome the surface-induced disorder, single-unit SRO in the form of (SRO)[1]-(STO)[5] superlattice geometry has been examined experimentally. However, different ground states of single-u.c. SRO superlattices are obtained, from a non-FM insulator [22,23] to FM insulator [35] to borderline FM metal [36]. They are in contrast to the theoretically suggested half-metallic state for 1 u.c. SRO layer confined within STO lattice [32].

In this paper, we report a combined experimental and theoretical investigation on $(SrTiO_3)^5$-$(SrRuO_3)^n$-$(SrTiO_3)^5$ ($STO^5$-$SRO^n$-$STO^5$) heterostructures ($n$ = 1, 2 u.c.). Electric- and magneto-transport measurements demonstrate that $STO^5$-$SRO^1$-$STO^5$ is insulating and non-FM, whereas $STO^5$-$SRO^2$-$STO^5$ is FM-metallic with a Curie temperature of ~128 K. Atomically resolved structural analysis reveals that octahedral tilts are absent in both heterostructures, thus ruling out such structural changes as a controlling factor for such drastic property differences. On the other hand, atomically resolved electron-energy-loss-spectroscopy (EELS) chemical maps show that $STO^5$-$SRO^1$-$STO^5$ is nonstoichiometric with substantial interface induced Ti-Ru intermixture, while $STO^5$-$SRO^2$-$STO^5$ is nearly stoichiometric. Density functional theory (DFT) calculations find that stoichiometric $STO^5$-$SRO^n$-$STO^5$ ($n$ = 1, 2 u.c.) are still FM-metallic without SRO octahedral tilts. Ru deficiency caused by Ti-Ru intermixing leads to the stabilization of AFM ordering and insulating behavior in the monolayer SRO indicating that the experimentally observed intermixing is indeed responsible for the observed loss of metallicity and ferromagnetism.

## II. EXPERIMENT

**Thin Film Growth:** Heterostructures of the form $STO^5$-$SRO^n$-$STO^5$ with $n$ = 1, 2 u.c. [see Fig. 1(a)] were fabricated via pulsed laser deposition (PLD) on $SrTiO_3$ (STO) substrates oriented with a (001) surface. Both $STO^5$-$SRO^1$-$STO^5$ and $STO^5$-$SRO^2$-$STO^5$ heterostructures have two repetitions of SRO/STO building blocks. The STO substrates were first sonicated in deionized water and then treated for 30 seconds in buffered hydrogen fluoride, followed by annealing at 950° C in an oxygen atmosphere to produce atomically smooth surfaces. The SRO and STO films were grown at 650° C with an oxygen pressure of 100 mTorr and 10 mTorr, respectively. A KrF excimer laser ($\lambda$ = 248 nm) laser repetition with a rate of 10 Hz (SRO) and 5 Hz (STO), and energy of 300 mJ (SRO) and 260 mJ (STO) was used. Post deposition, the samples were cooled down at ~12°/min to room temperature in 100 mTorr oxygen. The film thickness was monitored by an in-situ reflection high-energy electron diffraction (RHEED). Figure 1(b) and 1(c) show *in situ* RHEED results. Time dependent RHEED oscillations show stabilized layer-by-layer film-growth mode throughout the deposition process. Moreover, the RHEED pattern of SRO and STO sublayers indicates an atomically smooth film surface.

**Electrical transport and magnetic property measurements:** Electron transport measurements were performed via a Quantum Design Physical Property measurement system in a four-probe configuration. The magnetoresistance (MR) was measured at different temperatures by applying an external magnetic field along film normal. The samples magnetization was studied by using a Quantum Design Superconducting Quantum Interference Device Reciprocating Sample Option. The magnetization as a function of temperature M(T) measurement was obtained via first cooling the samples down to 5 K under the 0.2 T field, and then while warming in presence of 0.01 T, the M(T) data was collected.

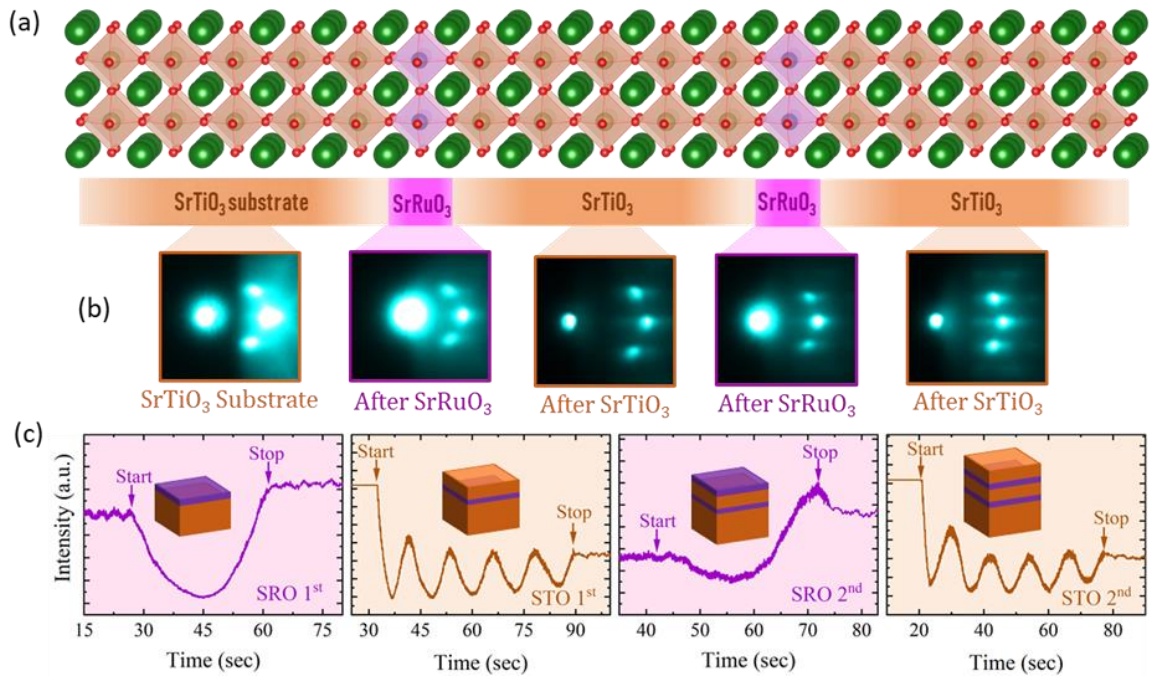

Fig. 1. STO$^5$-SRO$^1$-STO$^5$ heterostructure: (a) schematic sketch, and (b) *In-situ* RHEED paterns of STO substrate, SRO, and STO sublayers. (c) Time-dependent RHEED intensity profile.

**Scanning Transmission electron microscopy (STEM) and Electron energy loss spectroscopy (EELS):** STEM and EELS experiments were performed on a 200 kV JEOL ARM electron microscope at Brookhaven National Library equipped with double aberration correctors, a dual energy-loss spectrometer, and a cold field-emission source. TEM samples were prepared using a focused ion beam with Ga$^+$ ions followed by Ar$^+$ ions milling to a thickness of ~30 nm. The atomic-resolution STEM images were collected with a 21 mrad convergent angle (30 μm condenser aperture) and a collection angle of 67 – 275 mrad for high-angle annular dark-field (HAADF) and 11 – 23 mrad for annular bright-field (ABF) imaging. The atomic positions were obtained using two-dimensional Gaussian fitting following the maximum intensity. The microscope conditions were optimized for EELS acquisition with a probe size of 0.8 Å, a convergence semi-angle of 20 mrad, and a collection semi-angle of 88 mrad. Dual EELS mode was used to collect low-loss and core-loss spectra simultaneously for energy drift calibration in the collecting process. EELS mapping was obtained across the whole film with a step size of 0.2 Å and a dwell time of 0.05 s/pixel. The EELS background was subtracted using a power-law function, and multiple scattering was removed by a Fourier deconvolution method.

## III. RESULTS AND DISCUSSION

### A. Structure and composition

The structure and composition of the samples were investigated via atomic-resolution HAADF/ABF- STEM imaging and EELS mapping. The intensity in the HAADF image is roughly proportional to $Z^{\sim 2}$ (Z is an atomic number), depicting directly heavy-atom positions, whereas ABF imaging is useful for visualization of lighter atoms such as oxygen. The STEM images of $STO^5$-$SRO^1$-$STO^5$ with two single-u.c. SRO repeating blocks are shown in Fig. 2. The individual Sr (Z = 38), Ru (Z = 44), and Ti (Z = 22) atoms could be distinguished based on intensity contrast [see Figs. 2(a)-(d)], permitting us to determine the hetero-interfaces [see orange lines in Figs. 2(b)-(d)]. The HAADF-STEM images reveal that the crystalline lattice is coherent across the interfaces in the entire heterostructure. Furthermore, it can be seen from the HAADF image that the Ru column in the 1st SRO block is darker than that in the 2nd SRO block, which is related to severe Ti-Ru intermixing as will be discussed later.

To quantitatively examine the lattice-mismatch-induced structural distortions, we determined the out-of-plane (OOP) and in-plane (IP) lattice parameters from A-site atomic positions [see Figs. 2(e) and 2(f)]. The IP lattice parameter (*b*) of the SRO/STO interlayer is consistent with the STO substrate (Fig. 2f), indicating the sublayers are fully compressively strained (Bulk: $a_{STO}$ =3.905 Å and $a_{SRO}$ = 3.925 Å). The OOP lattice parameter (*c*) of the SRO blocks [Fig. 2(e)] is nearly the same as that of STO, suggesting a cubic symmetry, but the octahedral volume is smaller (59.85 ± 0.09 Å$^3$) than the bulk SRO (~ 60.37 Å$^3$) value of RuO$_6$, which is associated with Ti-Ru intermixture (discussed later).

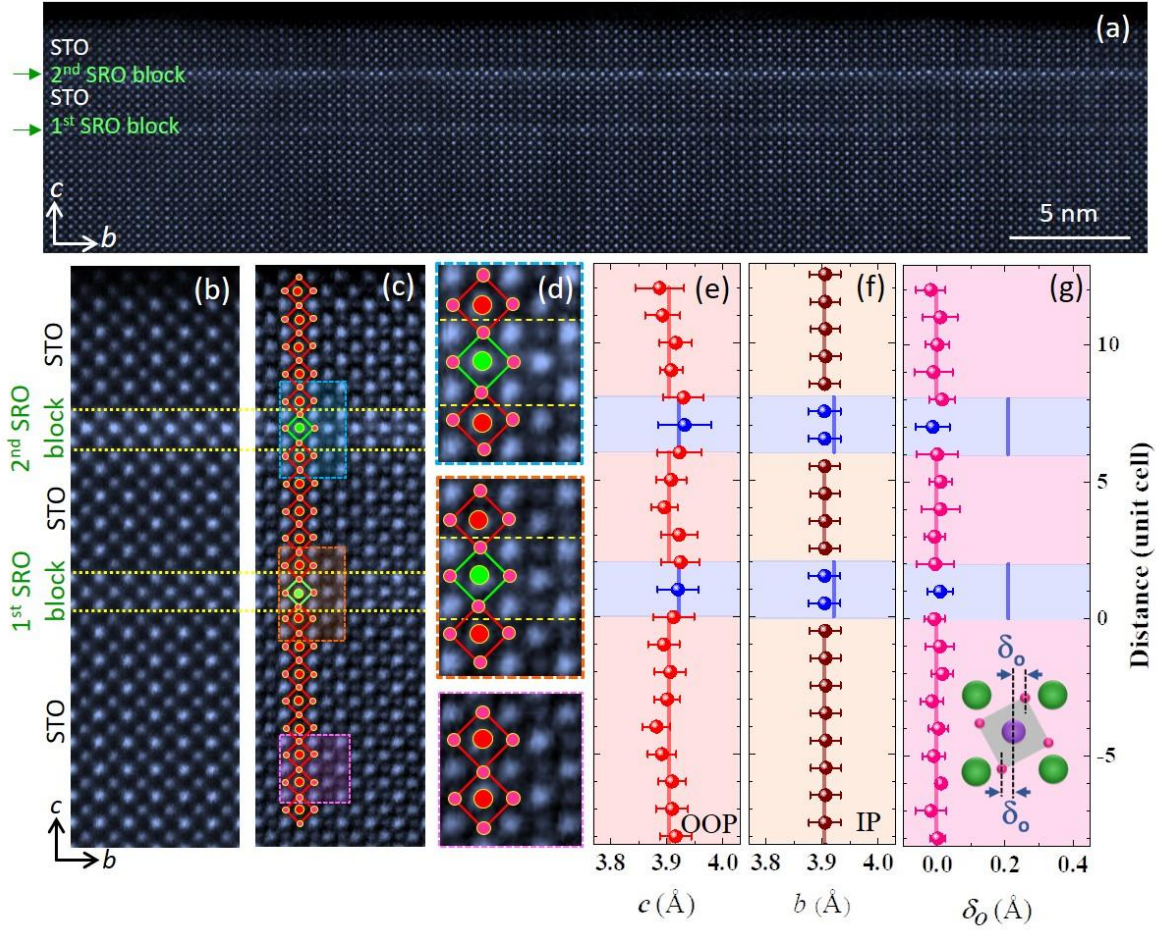

**Fig. 2.** Atomically resolved structure of STO$^5$-SRO$^1$-STO$^5$ (with two single-u.c. SRO repeating blocks) viewed along the [100] direction: (a) large area HAADF-, (b) high magnification HAADF-, (c) intensity-inversed ABF, and (d) zoomed inversed ABF-STEM images. The dotted orange line marks the interface, whereas red and green squares in panel (c) signify the projected octahedral shapes in STO and SRO. (e) out-of-plane and (f) in-plane lattice parameter $c$ and $b$, respectively, as a function of distance from the film-substrate interface. The error bar shows the standard deviations of the averaged measurements along the b-axis. The pseudocubic lattice parameters of bulk STO $(a_{STO} = 3.905\,\text{Å})$ and SRO $(a_{SRO} = 3.925\,\text{Å})$ are indicated by a red and blue solid line, respectively. (g) Variation of oxygen displacement $(\delta_O)$. The definition of $\delta_O$ is specified in the inset. The solid blue line in panel (g) marks the $\delta_O$ in bulk SRO (0.21Å). No octahedra tilts across the interfaces are observed.

Furthermore, in the ABF image [Figs. 2(c) and 2(d)] of the STO$^5$-SRO$^1$-STO$^5$ heterostructure, oxygen columns are visible hence permitting us to determine the octahedral geometry. We have determined the oxygen displacement ($\delta_O$) as a function of atomic distance (see Fig. 2g). The $\delta_O$ can represent octahedral distortion, as shown in the inset of Fig. 2g. The $\delta_O$ of SRO and STO interlayers are nearly constant and comparable to the value of bulk STO (Fig. 2g), suggesting that the $RuO_6$ octahedral tilt is completely suppressed, while the projected $RuO_6$ octahedra are illustrated with the red/green box in Figs. 2(c) and 2(d).

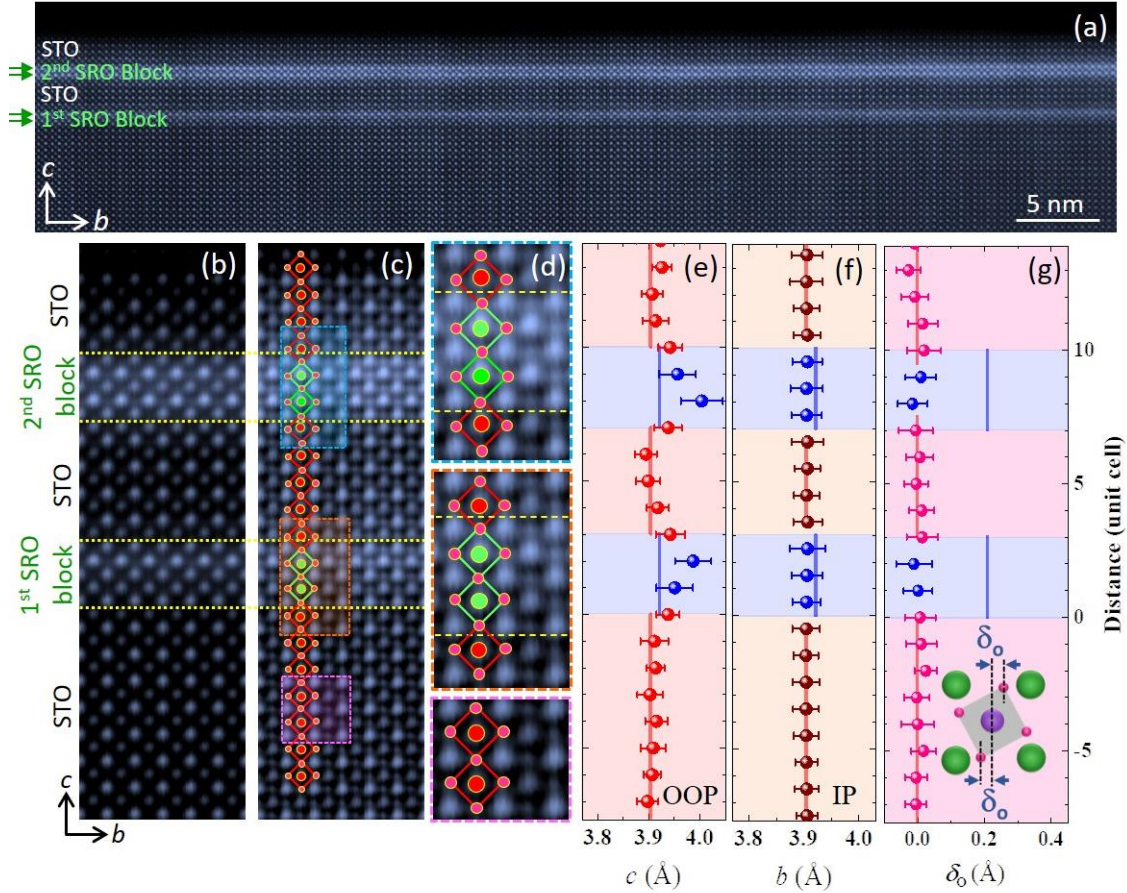

**Fig. 3.** Atomically resolved structure of $STO^5$-$SRO^2$-$STO^5$ (with two single-u.c. SRO repeating blocks) viewed along the [100] direction: (a) large area HAADF-, (b) high magnification HAADF-, (c) intensity-inversed ABF, and (d) zoomed inversed ABF-STEM images. The dotted orange line marks the interface, whereas red and green squares in panel (c) signify the projected octahedral shapes in STO and SRO. (e) out-of-plane and (f) in -plane lattice parameter $c$ and $b$, respectively, as a function of distance from the film-substrate interface. The error bar shows the standard deviations of the averaged measurements along the b-axis. The pseudocubic lattice parameters of bulk STO ($a_{STO}$ = 3.905 Å) and SRO ($a_{SRO}$ = 3.925 Å) are indicated by a red and blue solid line, respectively. (g) Variation of oxygen displacement ($\delta_O$). The definition of $\delta_O$ is specified in the inset. The solid blue line in panel (g) marks the $\delta_O$ in bulk SRO (0.21Å). No octahedra tilts across the interfaces are observed.

The atomic-level view of the STO$^5$-SRO$^2$-STO$^5$ heterostructure with two 2-u.c. SRO repeating blocks is shown in Fig. 3. The STEM images show coherent growth of the film where the two 2-u.c. SRO blocks are marked in Figs. 3(a)-(d). Additionally, the trend of having a nearly constant IP lattice parameter (*b*) reiterates that the SRO/STO interlayers are epitaxially strained [Fig. 3(f)], ensuring a high-quality heterostructure. The OOP lattice parameter (*c*) of the STO layer follows its bulk value [Fig. 3(e)], while the SRO blocks show increased OOP lattice parameter due to compressive strain, suggesting a tetragonal symmetry. This OPP lattice parameter (*c*) in 2-u.c. SRO blocks are also larger than that of monolayer SRO heterostructure [see Fig. 2(e)], which will be discussed later. On the other hand, oxygen displacement ($\delta_O$) as function of atomic distance is nearly zero across the interfaces throughout the heterostructure, advocating octahedral-tilt-angle suppression [see Figs. 3(c), 3(d) and 3(g)] Given that octahedral unit's preserve their connectivity via corner-shared oxygen atoms, the ultra-thin SRO confined between cubic STO blocks, facilitates the entire suppression of RuO$_6$ octahedral tilt angle (Ru-O-Ru $\cong$ 180°, cubic-like), leading to stabilization of the artificially engineered bond angle of SRO [37–39].

To determine the possible B-site cation (Ti, Ru) intermixing in the heterostructures, we took and analyzed STEM/EELS maps from many areas of the samples. The results are presented in Fig. 4. As shown in Figs. 4(a)-(c), the Ti atoms diffuse significantly into the Ru sites in STO$^5$-SRO$^1$-STO$^5$, especially the first SRO block (i.e., a single RuO$_2$ layer), resulting in the low Ru column intensity in the HAADF image [see Fig. 4(a)]. Quantitatively, the dopant Ti concentrations at the Ru sites were obtained from EELS profiles using the Lorentz-function-fitting method, with the Ti in the STO substrate as reference. We note a 61 ± 6 % Ti occupying the Ru site in the 1$^{st}$ (near substrate) SRO block and 25 ± 11% in the 2$^{nd}$ block of the STO$^5$-SRO$^1$-STO$^5$ heterostructure [see Figs. 4(b) and 4(c)]. On the other hand, the EELS mapping results from two representative areas of STO$^5$-SRO$^2$-STO$^5$ [one shown in Figs. 4(d)-(f) and the other in Figs. 4(g)-(i), respectively] are rather different from STO$^5$-SRO$^1$-STO$^5$. Though, the Ti concentration is still significant and varies from one region to another in the RuO$_2$ layer proximal to STO substrate of the 1$^{st}$ SRO block [~10% or less as in Fig. 4(f) to 60 % as in Fig. 4(i)]. Nevertheless, few Ti ions ($\leq$10 %) diffuse into the second RuO$_2$ layer of the 1$^{st}$ SRO block or both RuO$_2$ layers of 2$^{nd}$ SRO blocks [see Figs. 4(d)-(i)]. Overall, the spectroscopic observations suggest that the 2-u.c. SRO blocks in STO$^5$-

SRO$^2$-STO$^5$ are considerably more stoichiometric, while the single-u.c. SRO blocks in STO$^5$-SRO$^1$-STO$^5$ hold a higher Ru-cation off-stoichiometry due to interface intermixture. The substantial Ti atoms in the single-u.c. SRO blocks of the STO$^5$-SRO$^1$-STO$^5$ heterostructure bring the OOP lattice parameter (*c*) of SRO close to that of STO [Fig. 2(e)]. Additionally, the RuO$_6$ octahedral volume is reduced (59.85 ± 0.09 Å$^3$) in comparison to the bulk SRO (~ 60.37 Å$^3$) (STO bulk volume: 59.45 Å$^3$). In contrast, owing to nearly stoichiometric Ru concentration, the SRO blocks in STO$^5$-SRO$^2$-STO$^5$ retain their bulk-like pseudocubic unit-cell nature. Because of the in-plane compressive strain, the RuO$_6$ in STO$^5$- SRO$^2$-STO$^5$ shows slight elongation in the *c*-axis [see Fig. 3(e)] while maintaining RuO$_6$ octahedral volume (60.45 ± 0.19 Å$^3$) close to that in bulk SRO (~ 60.37 Å$^3$). In ultrathin oxide ABO$_3$ heterostructures, the interface B-site intermixture is unavoidable, irrespective of the growth method [40]. In fact, sensitivity of ruthenates to B-site disorder and volatile nature of Ru, makes it challenging to attain stoichiometric Ru films in the ultra-thin limit [41–43].

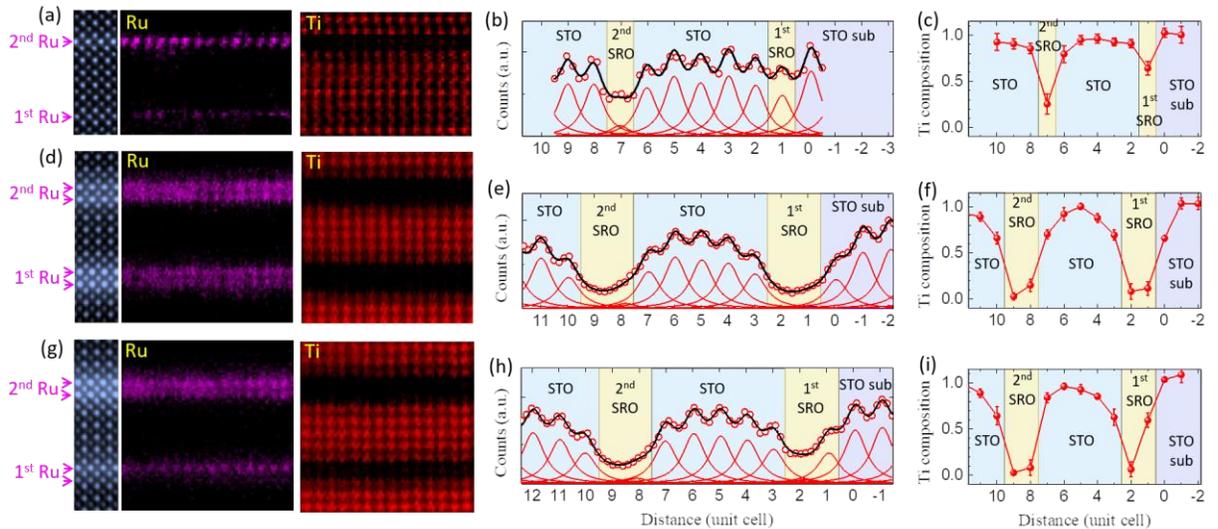

**Fig. 3.** STEM/EELS maps and elemental concentration: (a) HAADF images and EELS elemental maps from Ti-L$_{2,3}$, and Ru-M$_{2,3}$ edges, (b) Least-squares fit of Ti EELS intensity profiles (red dots) from averaging the Ti maps over the horizontal direction in (a), and (c) Ti composition profiles across STO$^5$-SRO$^1$-STO$^5$ heterostructure. The black curve is a sum of the Lorentzian peaks fixed at the Ti lattice sites. The Lorentzian peaks indicate Ti concentration. The same arrangement of STEM/EELS characterization for two representative mapping areas of STO$^5$-SRO$^2$-STO$^5$ heterostructure, (d-f) and (g-i), respectively. The distance (in u.c.) is defined with respect to the film-substrate interface.

## B. Transport & magnetic properties

After a thorough understanding of the structure and composition, we proceed to investigate the transport and magneto-transport properties. Figure 5 shows sheet resistance as a function of the temperature of $STO^5$-$SRO^n$-$STO^5$ ($n$ = 1, 2 u.c.) heterostructures. An insulating behavior of increasing resistance with lowering temperature is observed in $STO^5$-$SRO^1$-$STO^5$. Moreover, as shown in the inset of Fig. 5, the transport of $STO^5$-$SRO^1$-$STO^5$ can be fitted nicely with Efros-Shklovskii variable-range hopping model [39,40], where conductivity follows: $\sigma(T) = \sigma_0 e^{(T_{ES}/T)^{1/2}}$ ( $T_{ES} = \beta_{ES} e^2/\varepsilon k_B \xi$ is a characteristic temperature, $\xi$ is localization length, $e$ elementary charge, $\varepsilon$ dielectric constant. The linear fitting of data yields $T_{ES} \approx 30$ K), signifying the disorder-induced strong localization due to the Ti-Ru intermixture is likely the primary driving force for the insulating behavior. In the presence of strong localization effects, the kinetic theory of conductivity ($\sigma = e^2 k_F l/h$, where $k_F$ is Fermi wave vector, and $l$ is mean free path) breaks down, since the electronic mean free path ($l$) turn out to be equal to lattice spacing ($l \approx a$), leading to $k_F l \approx 1$, and system crosses the minimum metallic conductivity known as the Ioffe-Regel limit [44,45], where the sheet resistance limit is: $R_s = h/e^2 \approx 25\ k\Omega/\square$. Figure 5 confirms that the $STO^5$-$SRO^1$-$STO^5$ heterostructure sheet resistance exceeds the Ioffe-Regel limit in the measured temperature range. On other hand, the $STO^5$-$SRO^2$-$STO^5$ heterostructure shows the metallic character (see Fig. 5) in the measured temperature range. Though, the small resistivity upturn below ~ 25 K is observed, mainly caused by the disorder-induced weak localization effects [9,27,46].

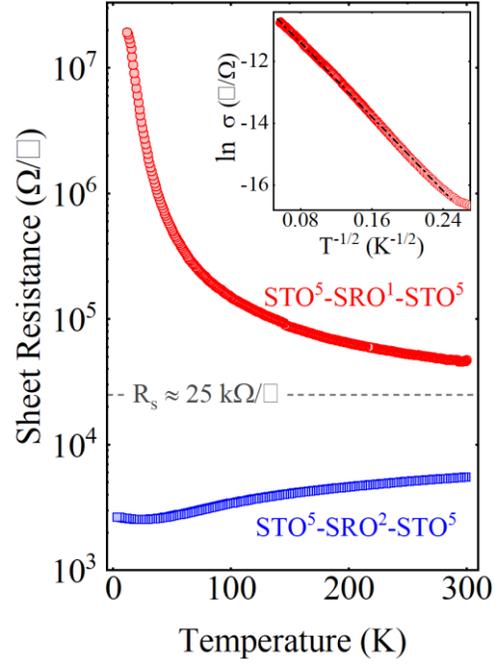

**Fig. 5.** Temperature-dependent sheet resistance of $STO^5$-$SRO^n$-$STO^5$ with $n$ =1, 2. The inset shows a logarithmic plot of $STO^5$-$SRO^1$-$STO^5$ conductivity versus $T^{-1/2}$, representing the Efros-Shklovskii variable range hopping model. The grey dashed line is the quantum of resistance ($R_s \approx 25$k $\Omega/\square$).

We have performed magnetoresistance (MR) measurements; MR = $\{\rho(H) - \rho(0)\}/\rho(0)$, here $\rho(H)$ and $\rho(0)$ are resistivities in the presence and absence of a magnetic field, respectively (the external magnetic field *(H)* is perpendicular to the film plane). The MR of STO[5]-SRO[1]-STO[5] shows a parabolic nature [Fig. 6(a)], signifying an absence of FM order. In contrast, the MR of STO[5]-SRO[2]-STO[5] shows a butterfly loop MR, representing the FM state [see Fig. 6(b)]. The maximum (~ 2 T) in the MR at 5 K hysteretic curve corresponds to coercive field $H_c$, whereas the point of forward and backward sweeps overlapping is the saturation field (~6 T). The hysteretic MR characteristic of FM ordering persists up to ~55 K, though non-parabolic MR dependence continues up to ~120 K. For manganite's [47–50],

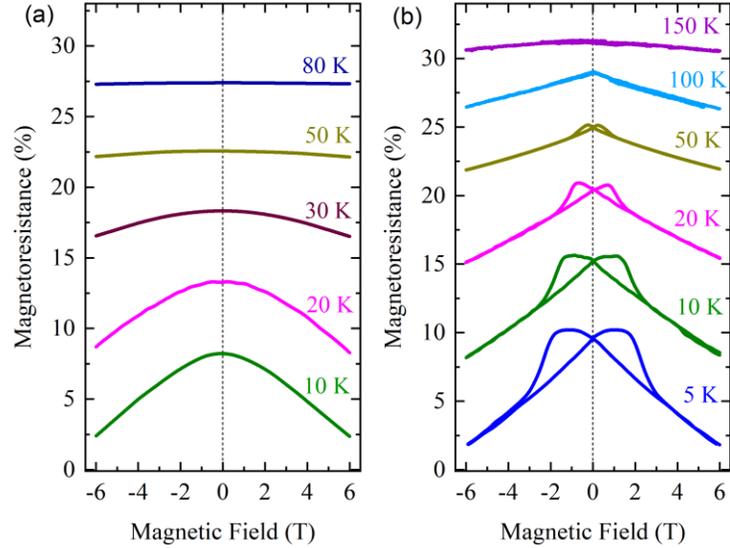

**Fig. 6.** Magnetoresistance of (a) STO[5]-SRO[1]-STO[5], and (b) STO[5]-SRO[2]-STO[5] measured at different temperatures.

granular magnetic systems [51,52], and SRO [53], the MR scales as MR $\propto [M(H)/M_s]^2$, while MR $\propto [\mu_0 H]^2$ for nonmagnetic-conducting systems [36,51,54], where $M_S$ is saturation magnetization and H the applied field. The magnetoresistance dependence on the magnetization (MR $\propto M^2$) suggests that electron transport depends on the magnetic moment's alignment within magnetic domains. As the magnetic moments are aligned, the carrier scattering decreases and so is the resistivity, while the resistivity is maximized at the coercive fields. Yet, trapping centers such as dislocations, defects, and non-magnetic inhomogeneity, might cause the pinning of domain walls, leading to enlarged switching fields [54]. Moreover, the disappearance of the butterfly loop before the actual $T_C$ is triggered by the dominance of thermal fluctuations over the pinning of domain walls [36,51,54]. Nonetheless, the presence of a butterfly-like feature in the MR below 55

K corroborates the presence of FM ordering in $STO^5$-$SRO^2$-$STO^5$, while the single-u.c. SRO is non-FM. The magnetization of $STO^5$-$SRO^n$-$STO^5$ ($n$ = 1, 2 u.c.) heterostructures was also studied via SQUID magnetometry. To measure magnetization as a function of temperature $M(T)$, the samples were first cooled down to 5 K under the 0.2 T field, and then while warming in presence of 0.01 T, the data was collected. The $STO^5$-$SRO^1$-$STO^5$ does not show any sign of FM transition, as the $M(T)$ curve is nearly flat (a) and the $M(H)$ is reminiscent of background hysteresis of the substrate [Figs. 7(a) and 7(b)], verifying the absence of FM state. However, $STO^5$-$SRO^2$-$STO^5$ exhibits robust FM ordering as shown by sharp paramagnetic to FM transition at $T_C$ ~ 128 K in $M(T)$ [Fig. 7(a)] and characteristic ferromagnetic hysteresis [see $M(H)$ in Fig. 7(b)]. Overall, the electron transport and magnetic properties results indicate that the presence of metallicity is important for stabilizing ferromagnetism since an insulating SRO is non-magnetic.

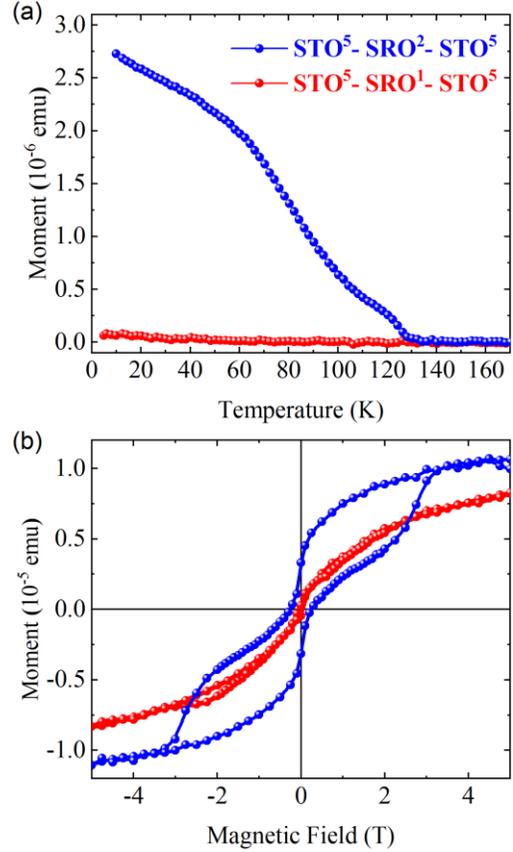

**Fig. 4.** $STO^5$-$SRO^n$-$STO^5$ [$n$ = 1, 2] (a) Temperature-dependent magnetization, and (b) magnetic hysteresis measured at 5 K.

## C. DFT Calculations of Intrinsic and Extrinsic Heterostructures

In order to better understand the magnetic and electronic nature of the present heterostructures, we have performed DFT calculations on $STO^5$-$SRO^n$-$STO^5$ ($n$ = 1,2 u.c.) heterostructures (see Appendix for computational details). The structural calculations reveal that both the $STO^5$-$SRO^1$-$STO^5$ and the $STO^5$-$SRO^2$-$STO^5$ heterostructures feature non-tilted octahedra in the SRO layers, in accord with the STEM data of Fig. 3. Thus, both experiments and theory rule out structural distortions to be the regulatory factor for the observed contrasting electro-magnetic properties in the two heterostructures.

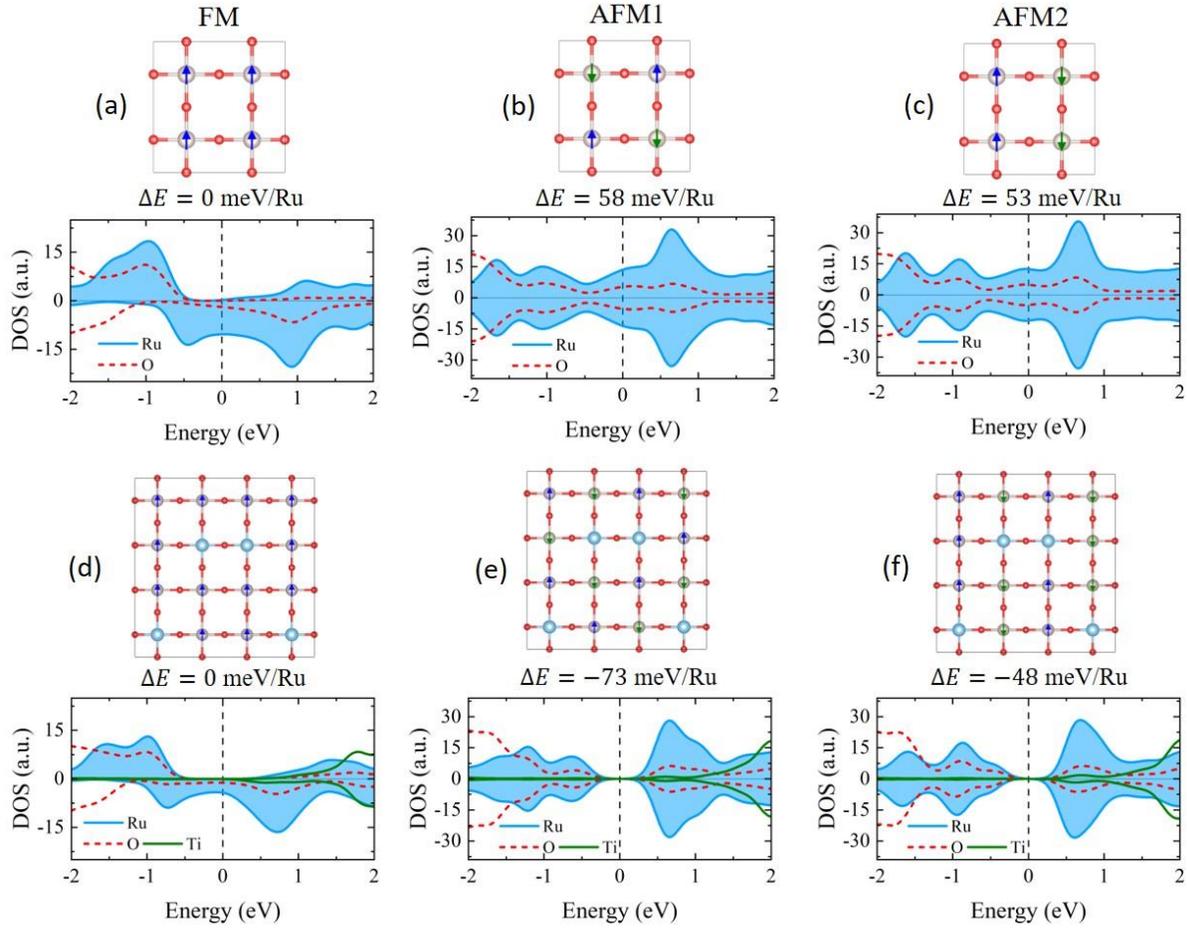

**Fig. 5.** Magnetic arrangements (top) and projected density of states (bottom) for the $RuO_2$ planes in $STO^5$-$SRO^1$-$STO^5$ heterostructures. Panels (a) through (c) are for a heterostructure containing a stoichiometric $RuO_2$ plane. Panels (d) through (f) are for 25% substitution of Ti for Ru within the $RuO_2$ plane. All densities of states are normalized to the same lateral system size so that they can be compared.

Beginning with the $STO^5$-$SRO^1$-$STO^5$ system, there are three possible magnetic arrangements for the Ru spins as shown in Fig. 8: (a) FM, (b) checkerboard AFM (AFM1), and (c) striped AFM (AFM2). The FM order is the ground state by 53 meV per Ru atom. In this FM state, the SRO layer in the $STO^5$-$SRO^1$-$STO^5$ system is half-metallic, namely metallic in only the minority-spin polarized electrons, while a gap appears between the spin-up $t_{2g}$ and $e_g$ states [Fig. 8(a)]. In contrast, both the higher-energy AFM phases are metallic as shown by the density of states in Figs. 8(b) and 8(c). The FM metallic ground state results are consistent with prior literature [32,33] and indicate that the observed non-FM insulating state is likely due to extrinsic effects and is not an intrinsic property of the system [28].

As demonstrated by the STEM results in Fig. 2 and Figs. 4(a)-(c) of $STO^5$-$SRO^1$-$STO^5$ heterostructure, the 1st SRO and 2nd SRO blocks contain 61 ± 6 % to 25 ± 11% Ru deficiency by Ti-for-Ru substitution, respectively. Therefore, from the original $STO^5$-$SRO^1$-$STO^5$ heterostructure model, we constructed a lateral superlattice containing 16 Ru sites and replace four of them with Ti to achieve 25% Ru deficiency via substitution. The chosen arrangement shown in Figs. 4(d) and 4(e) allows consideration of the same basic FM and AFM arrangements as before. We find that the introduction of the Ti leads to the energetic stabilization of both AFM arrangements relative to the FM solution. Furthermore, as seen in the density of states plots of Figs. 8(d)-(f), the introduction of Ti within the $RuO_2$ layer leads to shifts in the oxygen/transition-metal hybridization and opens a bandgap of 0.4 eV for both AFM orderings. Thus, we can conclude that the introduction of Ti within the $RuO_2$ layer via interfacial intermixture observed in STEM is responsible for the observed magnetic and electronic properties shown in Figs. 5(a), 6(a), and 7(a). This result complements the report by Boschker and coworkers [36], where using adsorption-controlled molecular-beam epitaxy, the SRO in monolayer limit is unveiled to hold metallicity and FM order, echoing the importance of the role of interface-induced intermixing and impurities in determining heterostructure properties [40,42,43,55]. Although other Ti-for-Ru substitutional patterns are possible, these calculations become prohibitively expensive with current methods. These results complements the report by Boschker and coworkers [36], where if defects are eradicated, a metallic-FM monolayer SRO could be stabilized.

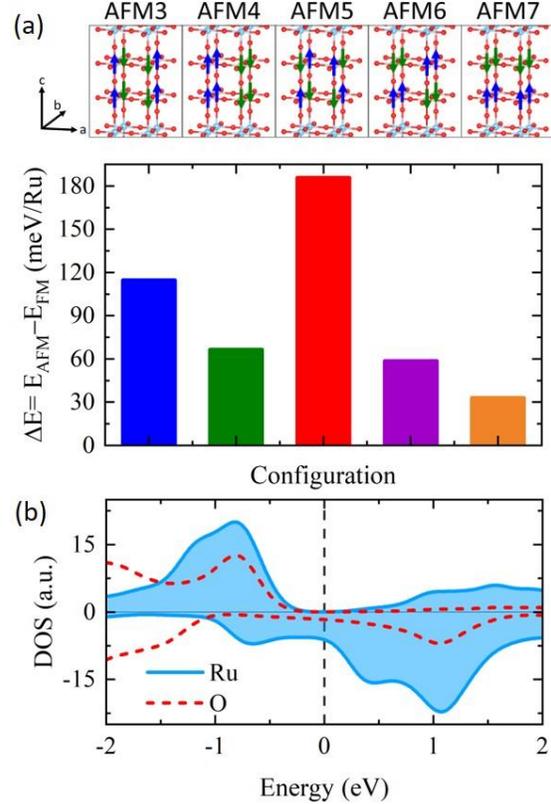

**Fig. 9.** (a) Possible AFM arrangements for the Ru sublattice in the $STO^5$-$SRO^2$-$STO^5$ heterostructure and their relative energy per Ru atom compared to the FM ground state. (b) Density of states for the lower $RuO_2$ plane in the $STO^5$-$SRO^2$-$STO^5$ heterostructure.

For the STO$^5$-SRO$^2$-STO$^5$ heterostructure, we consider four possible AFM arrangements [see the top of Fig. 9(a)] based on the prior single layer AFM arrangements (with differing coupling along the c-axis of the heterostructure) as well as an additional arrangement (AFM7) which has FM ordering within each plane. For AFM3 through AFM6, the total relative energy per Ru atom is much higher than the FM order compared to the stoichiometric STO$^5$-SRO$^1$-STO$^5$ case. The AFM7 case is also found to be higher in energy than the FM ordering, but is lower than the other STO$^5$-SRO$^2$-STO$^5$ energies since only the interlayer exchange coupling energy plays a role in the ΔE. As expected, the density of states for the layers (the lower layer is shown in Fig. 9(b), the upper layer looks similar) remains metallic like the stoichiometric STO$^5$-SRO$^1$-STO$^5$ structure and consistent with the experimental results shown in Figs. 6(b), 6(b), and 7(b).

## IV. CONCLUSIONS

In conclusion, by fabricating artificial heterostructures of the form STO$^5$-SRO$^n$-STO$^5$ ($n$ = 1, 2 u.c.), we have shown that the heterostructure with 2-u.c. SRO is metallic and FM ($T_C$ ~ 128 K) while the heterostructure with single-u.c. SRO is insulating and non-FM. There is no fundamental change in lattice structure with reducing the thickness of SRO, thus excluding structural modification as a controlling factor for such drastic property transitions. DFT results further suggest that a stoichiometric single-u.c. STO$^5$-SRO$^1$-STO$^5$ heterostructure would be FM and metallic. However, we observed that SRO in STO$^5$-SRO$^1$-STO$^5$ is non-stoichiometric, exhibiting a much greater amount of Ti in the SRO blocks due to Ti-Ru intermixture than in the STO$^5$-SRO$^2$-STO$^5$ heterostructure. The existence of these non-magnetic Ti impurities in single-u.c. SRO drastically affects the electronic structure as well as the coherence for FM ordering. Therefore, we conclude that it is the off-stoichiometry dictated by the Ti-Ru intermixture that leads to the insulating and non-FM behavior of SRO in the single u.c. thickness limit. The experimental data of Ref. [32,33,36] that find the single-unit-cell SRO film to be FM confirms our overall conclusions about both STO$^5$-SRO$^1$-STO$^5$ and STO$^5$-SRO$^2$-STO$^5$ heterostructures.

# ACKNOWLEDGMENTS

This work is primarily supported by the US Department of Energy (DOE) under Grant No. DOE DE-SC0002136. The electron microscopy work done at Brookhaven National Laboratory (BNL) was sponsored by the US DOE-BES, Materials Sciences and Engineering Division, under Contract No. DE-SC0012704. The use of BNL's Center for Functional Nanomaterials supported by BES Office of User Science Facilities for TEM sample preparation is also acknowledged. Work at Vanderbilt was supported by the U.S. Department of Energy, Office of Science, Division of Materials Science and Engineering under Grant No. DE-FG02-09ER46554 and by the McMinn Endowment at Vanderbilt University. Calculations were performed at the National Energy Research Scientific Computing Center, a DOE Office of Science User Facility supported by the Office of Science of the U.S. Department of Energy under Contract No. DE-AC02-05CH11231 as well as with resources provided by the Department of Defense's High-Performance Computing Modernization Program (HPCMP).

## Appendix A: COMPUTATIONAL DETAILS

The Vienna ab initio Simulation Package (VASP) [56] was used to perform density-functional-theory calculations utilizing the projector-augmented wave (PAW) method to describe core-valence electron interactions [57,58]. The Perdew-Burke-Ernzerhof generalized gradient approximation [59] with a Hubbard U correction (discussed in detail below) [60] was employed for the exchange-correlation functional. For the heterostructure calculations, a plane-wave basis cutoff energy of 450 eV was used and the Brillouin zone was sampled using a $4 \times 4 \times 1$ Γ-centered k-point grid. The heterostructures considered were terminated with SrO on both sides and consisted of 5-layers of $TiO_2$ on each side of the inserted $RuO_2$ layers. Each layer consists of a $2 \times 2$ lateral supercell of the cubic cell (4 Ru sites). For the consideration of 25% Ti substitution for Ru, a $2 \times 2$ lateral supercell of the previous structure was used (16 Ru sites before substitution). A vacuum spacing of 15 Å was used to prevent spurious interactions due to periodic boundary conditions along the z-axis. All structures were relaxed until the atomic forces were converged to less than 0.02 eV/Å.

The electronic structure of bulk SrRuO$_3$ (SRO) according to Rondinelli et al. [28] is best described by including electron-electron correlations in the form of a Hubbard term with U = 0.6 eV. For Hubbard U values > 2 eV, in bulk, SRO becomes half-metallic [28,61,62]. For SRO ultra-thins films, theoretical calculations even with U > 3 eV are incapable to reproduce the observed insulating-nonmagnetic state [28]. Due to these observations, Rondinelli et al. [28] concluded that the insulating-nonmagnetic state is not caused by intrinsic changes such as enhanced electron interactions or structural changes, but rather triggered by surface roughness, defects, or disorder. However, according to Verissimo et al. [32], the single layer of Ru confined between STO lattice is a minority-spin half-metallic ferromagnet with U = 4 eV. Furthermore, recent calculations using a Hubbard U of 3.5 eV produced the most reasonable results for few-layer SRO films sandwiched between BaTiO$_3$ in thin-film heterostructures [18]. Therefore, we adopted Hubbard U = 3.5 eV in the present study of the STO$^5$-SRO$^n$-STO$^5$ heterostructures. Variation of the effective Hubbard U by ±0.5 eV does not change the qualitative results when tested on the stoichiometric STO$^5$-SRO$^1$-STO$^5$ heterostructure (i.e., it remains a ferromagnetic metal).